\documentclass[prd,superscriptaddress,nofootinbib,showpacs]{revtex4}
\usepackage{graphicx}
\usepackage{dcolumn}
\usepackage{slashbox,multirow}
\usepackage{amsmath,amsthm,amssymb}

\begin{document}

\title{Late-time evolution of the gravitating Skyrmion}

\author{Stanis\l{}aw Zaj\c{a}c}

\affiliation{H.Niewodnicza\'nski Institute of Nuclear Physics, Polish Academy of Sciences,\\
ul. Radzikowskiego 152, 31-342 Krak\'ow, Poland}

%
%
\begin{abstract}
We study the dynamics of spherically symmetric solutions in the Einstein-Skyrme model. We focus
our attention on generic long-time evolution of initial data resulting in the formation of the
$B=1$ soliton, which plays the role of an attractor. We demonstrate that similarly to the case of
flat space evolution, the relaxation to the regular soliton (which we will call Skyrmion) is
universal and may be treated as a superposition of two effects -- quasinormal oscillations
responsible for intermediate asymptotics and a power-law tail describing the behavior of the
system at very long times. We determine the values of parameters describing asymptotics and
examine their dependence on the value of dimensionless coupling constant of the model.
\end{abstract}


\maketitle

\section{Introduction}

This paper is concerned with intermediate and final asymptotics of spherically symmetric
solutions in Einstein-Skyrme (ES) model and is meant as a natural extension of similar analysis
done in \cite{BChR1}. ES model is known \cite{BCh0, DHS} as the model which admits a surprisingly
rich spectrum of asymptotically flat static spherically symmetric solutions, both with and
without horizon. In particular, the static $B=1$ spherically symmetric solution (Skyrmion) is
stable \cite{HDS, BCh0, HSZ}. Such stable static solutions are very important configurations  in
any physical model as they are natural candidates  for endstates of a generic evolution.  During
this evolution the system relaxes to the final stable configuration radiating energy to infinity.
This radiation contains information about the nature of the central object, therefore the
understanding of the mechanism of the emission is of crucial importance.

 Skyrme model was suggested by T.H.R. Skyrme \cite{Skyrme} as a model
describing properties of baryons in particle physics. It is a nonlinear model of a chiral field
in which baryons are identified as solitons -- topologically stable static field configurations.
Apart from its usefulness in particle physics the model is interesting theoretically because
despite of its simplicity we may observe here -- both in flat space as well as in
self-gravitating case -- all features of the relaxation process. So for generic initial data
after short period of evolution sensitive to the initial conditions, the system reaches
intermediate asymptotics which may be described by the sum of exponentially decaying oscillations
(so called quasinormal modes) superimposed on final, static Skyrmion configuration. Its long-time
asymptotics is governed by the power-law tail.

The rest of this paper is organized as follows. We start in Section II with the specification of the model.
 Then in Section III we discuss the linear stability of the Skyrmion. Next Section IV
 is devoted to the calculation of parameters of quasinormal modes.
 In Section V we describe the numerical method used in calculation of the time evolution
 and demonstrate the results of calculation of quasinormal modes obtained directly from
  the numerical evolution as well as from the method described in Section IV. In this Section  we discuss also the results for power-law tails.

\section{Settings}
We consider the Einstein-Skyrme model, so the matter in our model is a chiral field -- an $SU(2)$ -- valued scalar function $U(x)$ with dynamics given by the Lagrangian \cite{Skyrme}:
\begin{equation}\label{matter}
L=\frac{f^{2}}{4}Tr(\nabla_{a}\nabla^{a}U^{-1})+\frac{1}{32e^{2}}Tr[(\nabla_{a}U)U^{-1},(\nabla_{b}U)U^{-1}]^{2} - \frac{1}{16 \pi G} R,
\end{equation}
where $\nabla_{a}$ is the covariant derivative with respect to the spacetime metric, $G$ -- gravitational constant and $R$ -- scalar of curvature.
By choosing appropriate units of mass and length we set the values of coupling constants $f$ and $e$ to one.

We restrict ourselves to the case of spherical symmetry and use the polar time slicing and the areal radial coordinate so we may parameterize the metric as follows
\begin{equation}\label{metric}
ds^{2}=-e^{-2\delta(r,t)}N(r,t)dt^{2}+N^{-1}(r,t)dr^{2}+r^{2}d\Omega^{2},
\end{equation}
where $d\Omega^{2}$ is the standard metric on the unit 2-sphere. For the chiral field we apply the usual hedgehog ansatz $U=exp(i\overrightarrow{\sigma}\cdot\hat{r} F(r,t)),$ where $\overrightarrow{\sigma}$ is the vector of Pauli matrices and $ \hat{r}$ -- unit radial vector.

\newpage

We denote  $\frac{\partial}{\partial t}$ and $\frac{\partial}{\partial r}$ by overdots and primes  respectively and introduce an auxiliary variable $P=ue^{\delta}N^{-1}\dot{F}$, where $u=r^{2}+2\sin^{2}F$ and as a result obtain the following set of ES equations:
\begin{equation}\label{dotF}
\dot{F}=e^{-\delta} \ N \ \frac{P}{u},
\end{equation}
\begin{equation}\label{wave}
\dot{P}=(e^{-\delta}NuF')'+ \sin(2F) e^{-\delta}\left(N(\frac{P^{2}}{u^{2}}-F'^{2})-\frac{\sin^{2}F}{r^{2}}-1\right),
\end{equation}
\begin{equation} \label{momentum}
\dot{N}=-\frac{2\alpha}{r}e^{-\delta}N^{2}PF',
\end{equation}
\begin{equation}\label{delta}
\delta'=-\frac{\alpha u}{r}\left(\frac{P^{2}}{u^{2}}+F'^{2}\right),
\end{equation}
\begin{equation}\label{hamilton}
N'=\frac{1-N}{r}-\frac{\alpha}{r}\left(2 \sin^{2}F+\frac{\sin^{4}F}{r^{2}}+uN(\frac{P^{2}}{u^{2}}+F'^{2})\right).
\end{equation}
\begin{flushleft}
Here $\alpha=4\pi Gf^{2}$ is dimensionless coupling constant.
\par\end{flushleft}

We are interested in regular asymptotically flat solution of ES equations. To ensure the
regularity in the center, $ N(r)=1\ +\ O(r^{2})$, we impose the following boundary condition
$F(r,t) \sim r$ for $ r \rightarrow 0$. Asymptotic flatness of initial data $N(r,0) = 1 + O(1/r)$
requires that $F(r,0) = B \pi + O(1/r^2) $ at infinity, where the integer $B$, which we will call
baryon number, is equal to the topological degree of the chiral field. As long as no horizon
forms the baryon number of initial configuration is preserved during the evolution. Therefore the
asymptotic flatness condition breaks the initial value problems into infinitely many disjoint
topological sectors labeled by the baryon number $B$. Below we consider mainly $B=1$ sector. In
is a well established fact \cite{BCh0} that in this sector for $\alpha \leq \alpha_{crit} \simeq
0.040378$ the Eqs.(\ref{dotF}-\ref{hamilton}) have two regular static solutions. One of them is
linearly stable and is identical with flat Skyrmion for $\alpha = 0$, so we regard it just as a
gravitationally distorted Skyrmion. The second one, which we denote by $X^u$ is  unstable. This
two solutions coalesce at $\alpha=\alpha_{crit}$ and disappear for $\alpha > \alpha_{crit}$.

\section{Linear stability of gravitating Skyrmion}

We are interested in relaxation processes of solutions to the static configuration,
therefore we start with the linear stability analysis of Skyrmion.
As we restrict ourselves only to the spherically symmetric perturbations,
we use the following standard decomposition of solutions:
\begin{equation}\label{decom]}
F(r,t)=S(r)+f_{1}(r,t),\quad N(r,t)=N_{0}(r)+n_{1}(r,t),\quad\delta(r,t)=\delta_{0}(r)+\delta_{1}(r,t),
\end{equation}
where ($ S(r), N_{0}(r), \delta_{0}(r) $) denote static Skyrmion configuration. We plug this
decomposition into Eqs.(\ref{dotF}-\ref{hamilton}) and linearize. As it was demonstrated in
\cite{HDS}, the analysis of such system simplifies considerably because in spherical symmetry the
metric perturbations which enter the pulsation equation for $f_1(r,t)$ are completely determined
by the matter perturbations. Exploiting this and introducing auxiliary field $v(r,t)$ defined by
the formula
\begin{equation}\label{auxfield}
v(r,t)=\frac{f_{1}(r,t)}{\sqrt{r^{2}+\ 2 \sin^{2} S}},
\end{equation}
we obtain the linear pulsation equation for $v(r,t)$:
\begin{equation}\label{pulsation_v}
e^{\delta_{0}}N_{0}^{-1}\ddot{v}-\ (e^{-\delta_{0}} N_{0}v')'+\ e^{-\delta_{0}} U v=\ 0.
\end{equation}
Let us note that the potential $U=\ U_{0}+ \alpha U_{1}+ \alpha^{2} U_{2}$ is determined entirely by the functions describing the static Skyrmion and is given by the formulae:
\begin{equation}\label{pot0}
U_{0}=\frac{2}{r^{2}}-\ \frac{1}{(1+\ 2w^{2})^{2}}\left(4w^{2}(1+\ 3w^{2}+\ 3w^{4})+\ \frac{N_{0}-\ 1}{r^{2}}\right),
\end{equation}
\begin{equation}\label{pot1}
U_{1}=8w(1+w^{2}) S'\cos{S}-\ \frac{w^{2}(2+\ w^{2})}{1+\ 2w^{2}}-\ 2(1+\ 2w^{2})S'^{2},
\end{equation}
\begin{equation}\label{pot2}
U_{2}=2(1+\ w^{2})(2+\ w^{2}) S'^{2}\sin^{2}S.
\end{equation}
Here $w\equiv{\sin(\frac{S}{r}})$.

Let us examine the behavior of the finite part of potential $V(r) = U(r) - 2/r^2$ for large r. We
may easily find, that this behavior in the flat space and in the case of the self-gravitating
model are different. In both cases we have $S(r) \sim \pi - 1/r^2$ what means that $w \sim 1/r^3
$. In case of the flat model we have additionally $\alpha=0,\ N_{0}(r)=1$; this means that potential
$V(r) \sim 1/r^6 $. For self-gravitating case $N_{0} \sim 1 - b/r$ what leads to the relation
$V(r) \sim 1/r^3$ for $r \rightarrow 0$. This difference will be important in the discussion of
power-law tails.

Using the radial coordinate $\rho$ defined by
\begin{equation}\label{rho}
\frac{d\rho}{dr}=e^{\delta_{0}(r)-\delta_{0}(0)}N_{0}^{-1},\ \rho{(0)}=0,
\end{equation}
and assuming the time-dependence of the form $v(r,t)=e^{-ikt}\psi(\rho)$, we transfer Eq.(\ref{pulsation_v}) into a radial p-wave Schr\"odinger equation \cite{HDS}
\begin{equation}\label{schrodinger}
\left(-\frac{d^{2}}{d\rho^{2}}+\frac{2}{\rho^{2}}+\ V(\rho)\right)\psi=k^{2}e^{2\delta_{0}(0)}\psi.
\end{equation}
Here the potential
\begin{equation}
V(\rho)=e^{-2\delta_{0}(r)+2\delta_{0}(0)}U(r(\rho))-\ \frac{2}{\rho^2}.
\end{equation}
is bounded for all $\rho$. The investigation of linear stability is then reduced to study the
eigenvalue problem (\ref{schrodinger}) in the space of square integrable functions. This was done
in \cite{BCh0}, where it was found, that the spectrum around the Skyrmion is continuous and
positive, $k^2 > 0$. This means that the gravitating Skyrmion is linearly stable. If we make
similar calculation for $X^u$  we find out that Eq.(\ref{schrodinger}) has exactly one bound
state with $k_{b}^2 < 0$, indicating an instability -- exponentially growing mode with the
exponent $\gamma_b = i k_b > 0$. In addition, for $\alpha \rightarrow \alpha_{crit}$, the
instability exponent $\gamma \rightarrow 0$. Numerical calculation  done in \cite{HSZ, BCh1}
confirm the above results also in nonlinear analysis.

\section{Calculation of Quasinormal modes}

Linear stability of the Skyrmion is an important property of ES system because it tells us that
this solution is an attractor describing the final configurations in $B=1$ sector of initial data
problem. However it does not provide a direct information about the way this configuration is
reached in the evolution. To obtain the intermediate and long time asymptotics of solutions of
Eqs.(\ref{dotF}-\ref{hamilton}) one usually discuss the Eq.(\ref{schrodinger}) describing the
perturbation of Skyrmion. A problem of this type -- especially perturbations of black holes and
compact objects -- has a long history starting with the papers of Regge and Wheeler \cite{RW},
Zerilli \cite{Zerilli} and Price \cite{Price}, for a review see  \cite{LivingRev} and
\cite{Chandra}. As we will see the perturbations of Skyrmion at intermediate times are described
by quasinormal modes, whereas late time asymptotics is controlled by a power-law tail. To
simplify the notation we will neglect the difference between the radial coordinates $r$ and
$\rho$, so the Eq.(\ref{schrodinger}) takes the form:

\begin{equation}\label{radial}
   -\Psi'' + (\frac{2}{r^2} + V(r) ) \Psi = k^2 \Psi.
\end{equation}

 In the case of ES system by quasinormal mode we will understand the solution of Eq.(\ref{radial}) satisfying the outgoing wave conditions for $r \rightarrow \infty$:

\begin{equation}\label{outgoing}
   \Psi(r) \propto e^{ik\rho},  \ \ k=\omega - i \gamma, \ \ \gamma >0,
\end{equation}
and regular at the center.

We expect, that the mode with the least damping $\gamma$ will dominate the intermediate
asymptotics in the $B=1$ sector. To find a proper solution we have to apply the above boundary
condition to the solution of Eq.(\ref{radial}); this will quantize the eigenvalue $k$.

\newpage

However, it is not easy to implement the boundary condition of this type (especially for long--range
potentials). To see that let us remark, that for $r > R$, where $R$ -- the range of potential,
any regular solutions $\Psi$ of Eq.({\ref{radial}}) may be decomposed in terms of Riccati-Hankel
functions
 $\hat{h}_{1}^{(+)}(kr)$ and $\hat{h}_{1}^{(-)}(kr)$ as follows:
\begin{equation}\label{decomposition}
  \Psi =  a_{+}(k) \ \hat{h}_{1}^{(+)}(kr) + a_{-}(k) \ \hat{h}_{1}^{(-)}(kr).
\end{equation}
The Riccati-Hankel functions have the following asymptotic behavior: $\hat{h}_{1}^{\pm}(kr) \sim e^{\pm ikr}$, hence the  outgoing wave boundary conditions
correspond to the zeros of $a_{-}(k)$ coefficient. However, this is numerically difficult to achieve as the "unwanted" ingoing component of
(\ref{decomposition}) decreases exponentially with $r$. Therefore, we try to remove a numerically small ingoing component on the background of large
outgoing component. To achieve this we should have there a very good resolution, at least of order of $ O(e^{-2\gamma R })$. For this reason a usual procedure
of shooting from $r=0$ and varying $k$ until $a_{-}(k)$ vanishes (naive shooting) is not feasible, especially in case of $\gamma R \gg 1$. We have found
that much better way of proceeding is to use a shooting-to-a-fitting-point technique which is much more accurate and additionally may be generalized to the
case of long-range potentials. In this technique we shoot from two sides -- $r=0$ (requiring that the solution is regular for $r \rightarrow 0$) and some large $r_2$,
where the decomposition (\ref{decomposition}) works and try to match both solutions at some intermediate $r_f$.

Technically \cite{BChR1}, we use the amplitude--phase representation for the solution of Eq.(\ref{radial}): $\Psi=A \ exp(i \Phi)$. As a result, this equation for complex function $\Psi$ is replaced by the following system of two equations for real functions $A$ and $\Phi$:
\begin{equation}\label{amp_phase1}
A''-A\phi'^{2}-\left(\frac{2}{r^{2}}+V+\gamma^{2}-\omega^{2}\right)A=0,
\end{equation}
\begin{equation}\label{amp_phase2}
A\phi''+2A'\phi'-2\omega\gamma A=0.
\end{equation}
Regularity at the center requires:
\begin{equation}
A(r)\sim r^{2}\qquad  \text{and} \qquad \phi(r)\sim\frac{\omega \gamma}{5}r^{2} \qquad \text{for} \qquad r \rightarrow 0.
\end{equation}
We solve this equations numerically from $r=0$ up to some relatively small intermediate $r_f$.
For $r>r_f$ we replace the Schr\"{o}dinger equation (\ref{radial}) by its Riccati form
\begin{equation}\label{Riccati}
g'+g^{2}-2/r^{2}-V+k^{2}=0,
\end{equation}
where $g$ is defined as a logarithmic derivative of $\Psi$: $g=\frac{\Psi'}{\Psi}=\frac{A'}{A}+i\phi'$.
We solve this equation backwards in $r$ from some large $r_2$ to the intermediate $r_f$ defined above, starting with the initial value
  $g(r_2)=\frac{ k (\hat{h}_{1}^{(+)}(k r_2))'}{\hat{h}_{1}^{(+)}(k r_2)}$.
We assume, that $r_2$ is large enough, so the Riccati-Hankel function $\hat{h}_{1}^{(+)}(kr)=(-i+\frac{1}{kr})e^{ikr}$
which is exact outgoing solution of the free ($V=0$) Riccati equation approximates  well the solution of the full Riccati equation. Now the quantization condition for the quasinormal modes is the matching of logarithmic derivatives at intermediate matching point $r_f$.

\section{Numerics and Results}

We have performed extensive numerical studies of long time asymptotics of solutions in ES model.
To this end we have solved initial value problem (\ref{dotF}--\ref{hamilton}) for various baryon
numbers, coupling constant and initial data. To do this we apply the standard method of
lines to the pair of dynamic equations (\ref{dotF}) and (\ref{wave}). This means that we discretize space and replace spatial derivatives by proper algebraic difference approximations. In that way we change the original set of dynamical PDEs into a system of ODEs, which may be solved by standard methods. Usually we use 5-point, fourth-order accurate spatial discretization and solve the resulting ODE system by means of fourth-order Runge-Kutta method. In addition, on each time layer we update the metric functions $N$ and $\delta$ by solving the slicing condition (\ref{delta}) and hamiltonian constraint (\ref{hamilton}).  Here we also use fourth-order Runge-Kutta method with spline interpolation for the values of grid functions at the
positions out of the grid which are also required by the integration procedure. As a result we have finite difference method which is fourth-order accurate both in space and time. We apply this method to the calculation of time evolution which starts with different initial conditions. Typical examples of initial data are:

\begin{equation}\label{initialB1}
 F(t=0,r)= \tanh(x/s), \qquad   P(t=0,r)=0
\end{equation}
for $B=1$ sector and
\begin{equation}\label{initialB0}
 F(t=0,r)= A r^{3} \exp[-((r-r_0)/s)^4], \qquad P(t=0,r)=0
\end{equation}
for $B=0$ sector of initial value problem.

\newpage

 To ensure the regularity at the origin we impose the boundary conditions: $F(t,r=0)=0, P(t,r=0)=0$. On the outer boundary we apply outgoing wave boundary condition. This is not enough when we calculate the properties of tails; in order to avoid the contamination of results by the part of solution reflected from the outer boundary, we use the size of the grid big enough, so the calculation stops before the reflected signal reaches the observation point. In calculations of this kind the quadruple precision is crucial, otherwise the accumulation of round-off errors spoils the results at late times.

\begin{figure}[htbp]
\begin{tabular}{cc}
\includegraphics[width=0.45\textwidth]{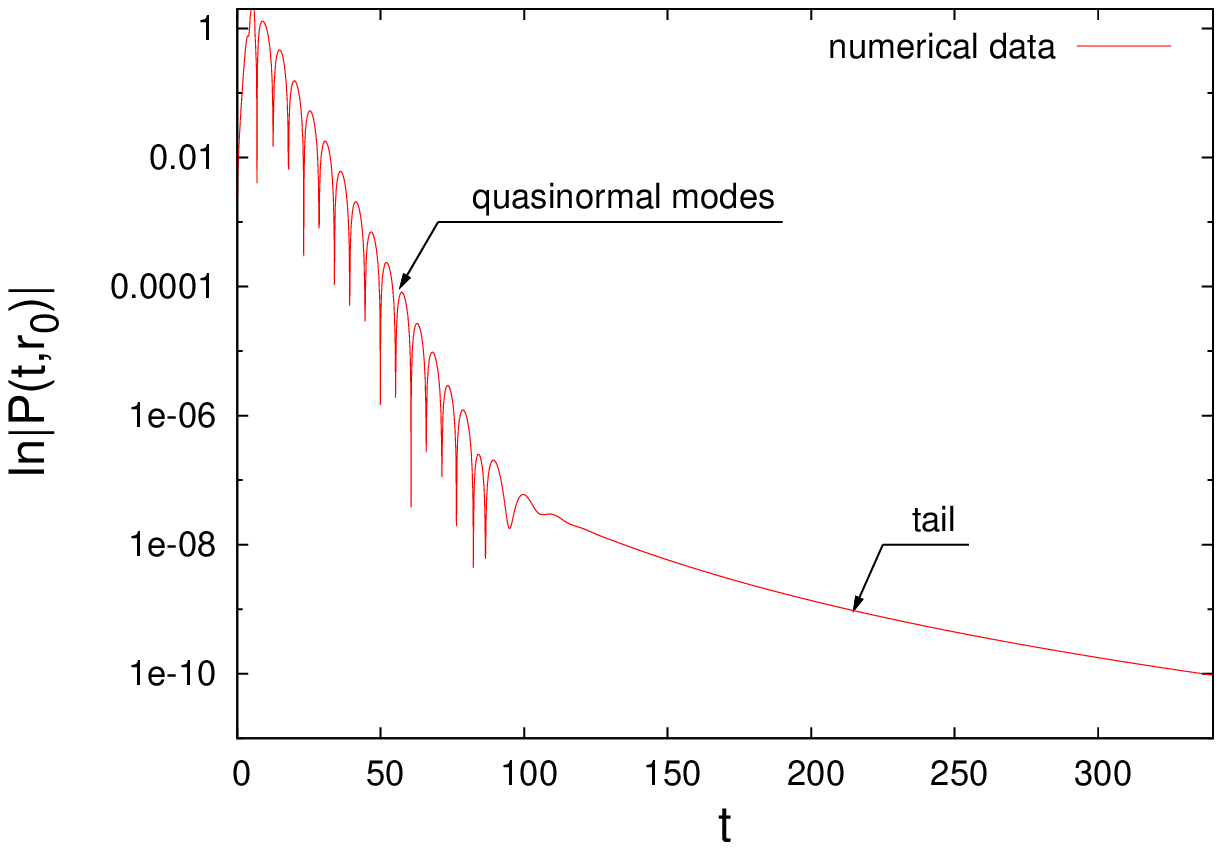}
&
\includegraphics[width=0.45\textwidth]{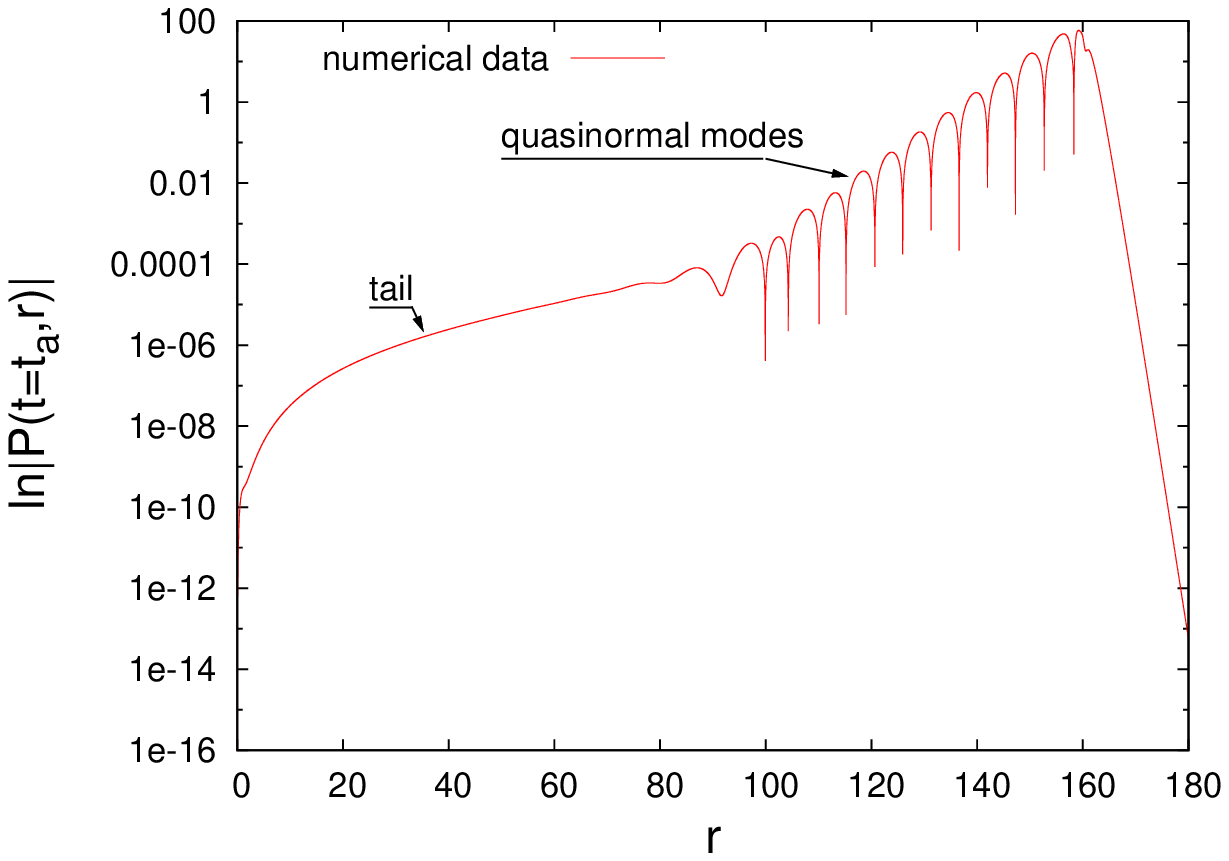}
\end{tabular}
\caption{An example of time evolution in $B=1$ sector. Left panel: we plot ln$|P(t,r_{0}=5)|$ as
a function of time and observe different phases of relaxation process. Right panel: the snapshot
of a solution for $t_a= 160$} \label{fig1}
\end{figure}

\begin{figure}[htbp]
\begin{center}
\includegraphics[width=0.45\textwidth]{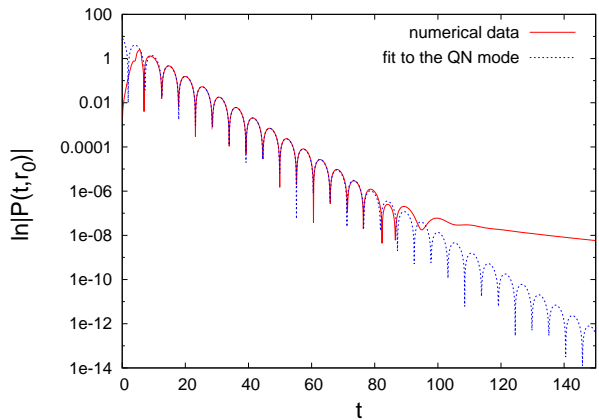}
\caption{Intermediate asymptotics of relaxation towards Skyrmion. Fitting the exponentially damped oscillation $P(t,r_{0})=Ae^{-\gamma t}|cos(\omega t+\alpha)|$ to the numerical data we may estimate the parameters of the least damped quasi-normal mode.}
\label{fig2}
\end{center}
\end{figure}

The typical results of long time evolution in $B=1$ sector is demonstrated in Fig.\ref{fig1}.a. In this figure a value of the field $P$  measured at a fixed value of $r$  is plotted as a function of $t$.
In this figure we may observed all features of relaxation process towards a stable Skyrmion  configuration: the beginning of evolution is strongly dependent on the shape of initial data. After some time an intermediate asymptotics sets in ($5 < t < 100$) where  the relaxation manifest itself as a damped oscillations. Finally, at large times ($t > 100 $) we observe the power-law tail. The same features may be observed on a complementary picture where a late-time snapshot of solution (see Fig.\ref{fig1}.b),
i.e. $|P(t=t_a,r)|$ as a function of $r$ is seen. Here we may also observed, that the fragment of solution corresponding to the quasi-normal modes treated as a function of $r$ grows exponentially, what makes the procedure of finding the mode parameters difficult numerically.
Using numerical data we may extract the parameters of least-damped quasi-normal mode (see Fig.\ref{fig2} for details).

To calculate the parameters of quasi-normal modes we use also the semi-analytic method described in Section IV.

\newpage

\begin{figure} [htbp]
\begin{tabular}{cc}
\includegraphics[width=0.45\textwidth]{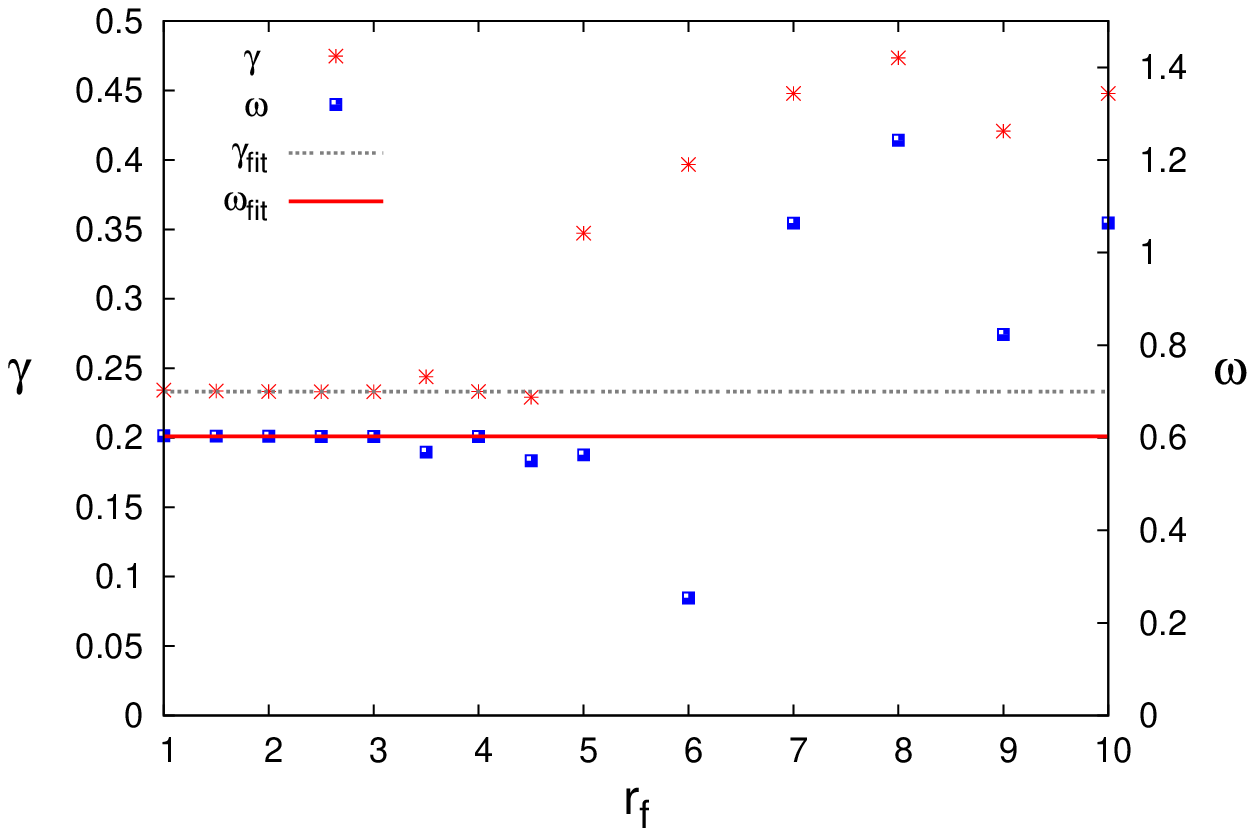}
&
\includegraphics[width=0.45\textwidth]{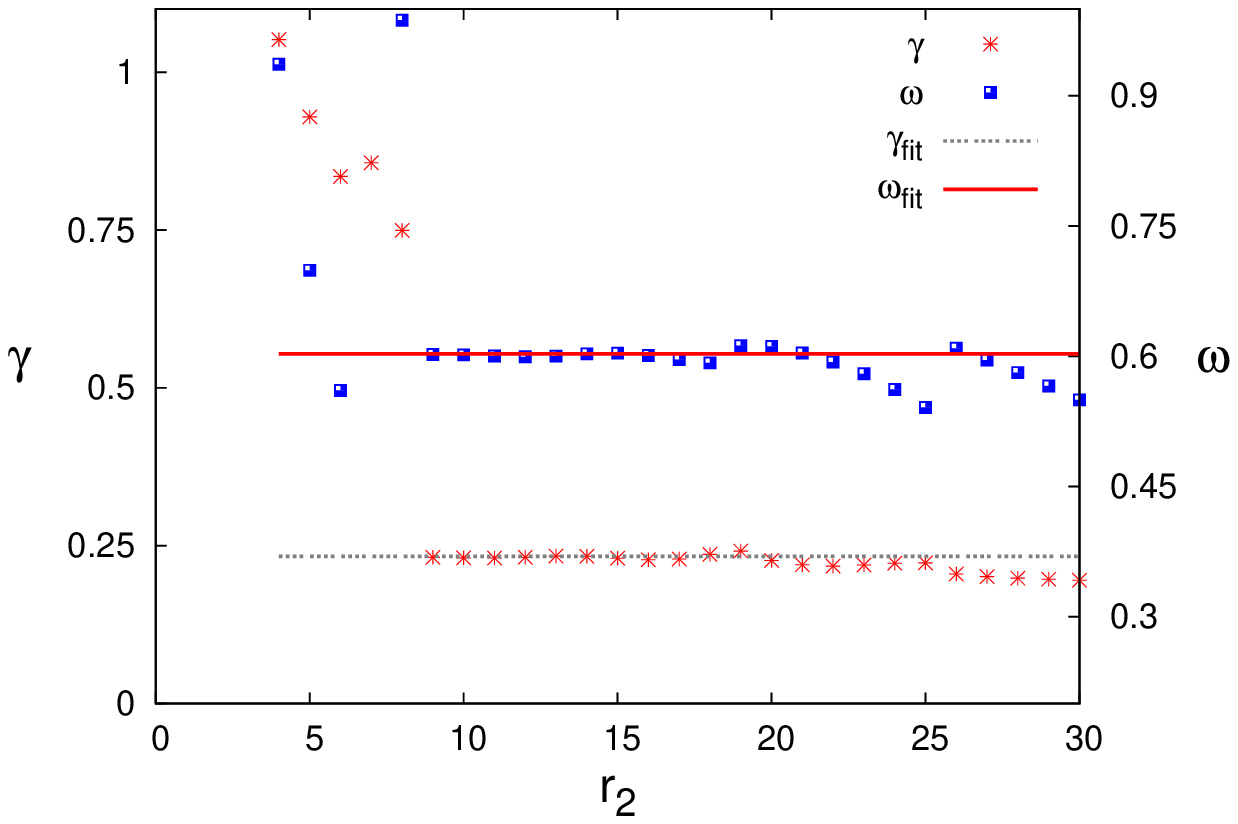}
\end{tabular}
\caption{\small{The sensitivity of results for the parameters ($\omega, \gamma$) obtained via
shooting-to-a-fitting-point technique on $r_f$ and $r_2$ parameters. Flat regions in both plots demonstrate
the robustness of the method.}} \label{fig4}
\end{figure}

To achieve this, we solve -- using standard fourth-order Runge-Kutta method --
Eqs.(\ref{amp_phase1}), (\ref{amp_phase2}), (\ref{Riccati}) together with time-independent version of
system (\ref{dotF}--\ref{hamilton}), which gives the potential $V$. To get the value of ($\omega,
\gamma$) parameters of the least damped quasi-normal mode we proceed as follow: we treat the
method described in Section IV as a function $T$ which transforms starting values of mode
parameters into the values corresponding to the mode in question:
\begin{equation}  T : (\omega_0, \gamma_0) \rightarrow (\omega, \gamma).
\end{equation}
We require that the true solution corresponds to a fixed point of transformation $T$ with
the universal values of starting mode parameters $\omega_0=1, \gamma_0=1$.
In addition we require that the $T$ procedure is not sensitive to the values of the intermediate matching point $r_f$ and external shooting point $r_2$. This means, that if we look at the value of the solution as a function of $r_f$ and $r_2$ we should observe a plateau. Fig.\ref{fig4} shows that it is really the case. We observe, that for $r_2=14$, the solution of semi-analytic method does not depend on the value of $r_f$ for  $1 < r_f < 3$. Similarly -- if we fix $r_f=2$
than the solution of semi-analytic method does not depend on the value of $r_2$ for $9  <  r_2 < 18$. In addition in both cases the value of semi-analytic solution agrees with the solution obtained as a fit to the numerical data.
This confirms that the semi-analytic method works well and is a good tool for estimation of the parameters of the least damped quasi-normal mode.

\begin{figure}[htbp]
\begin{tabular}{cc}
\includegraphics[width=0.45\textwidth]{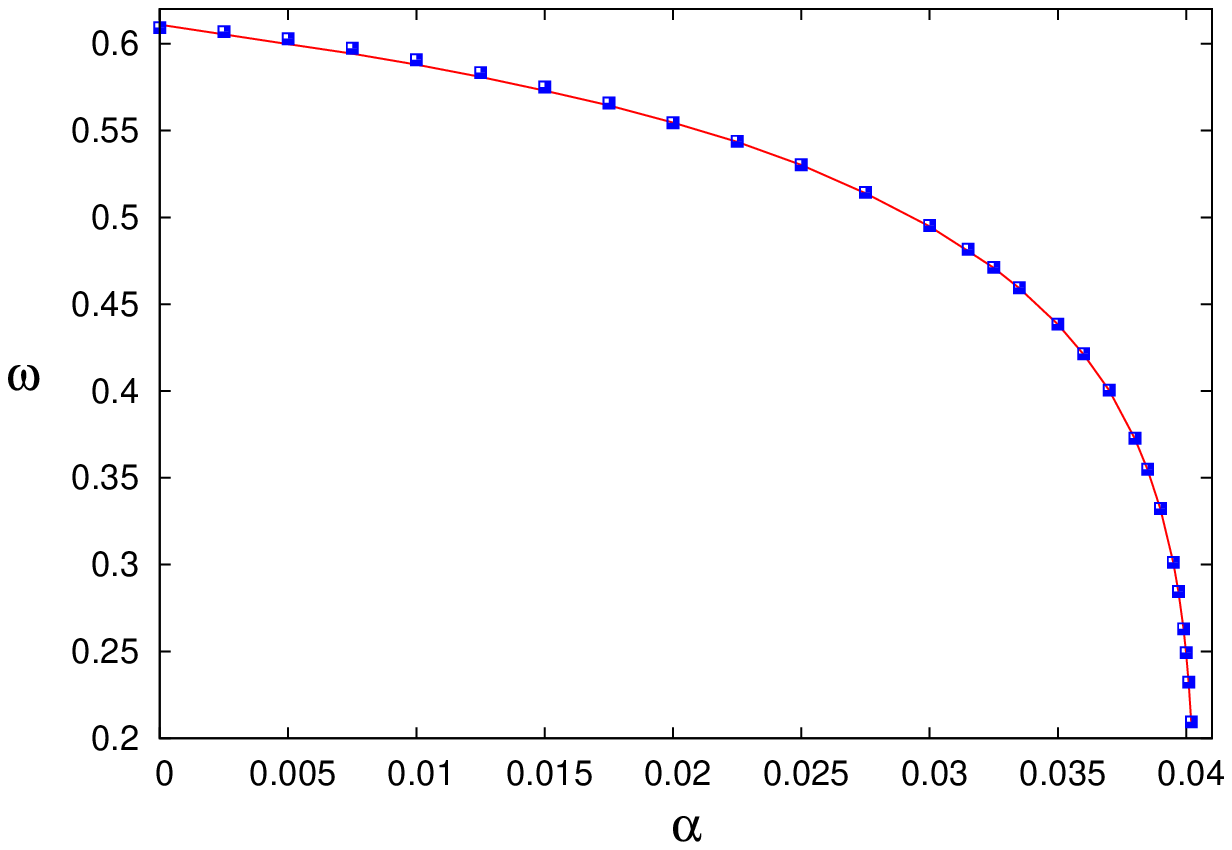}
&
\includegraphics[width=0.45\textwidth]{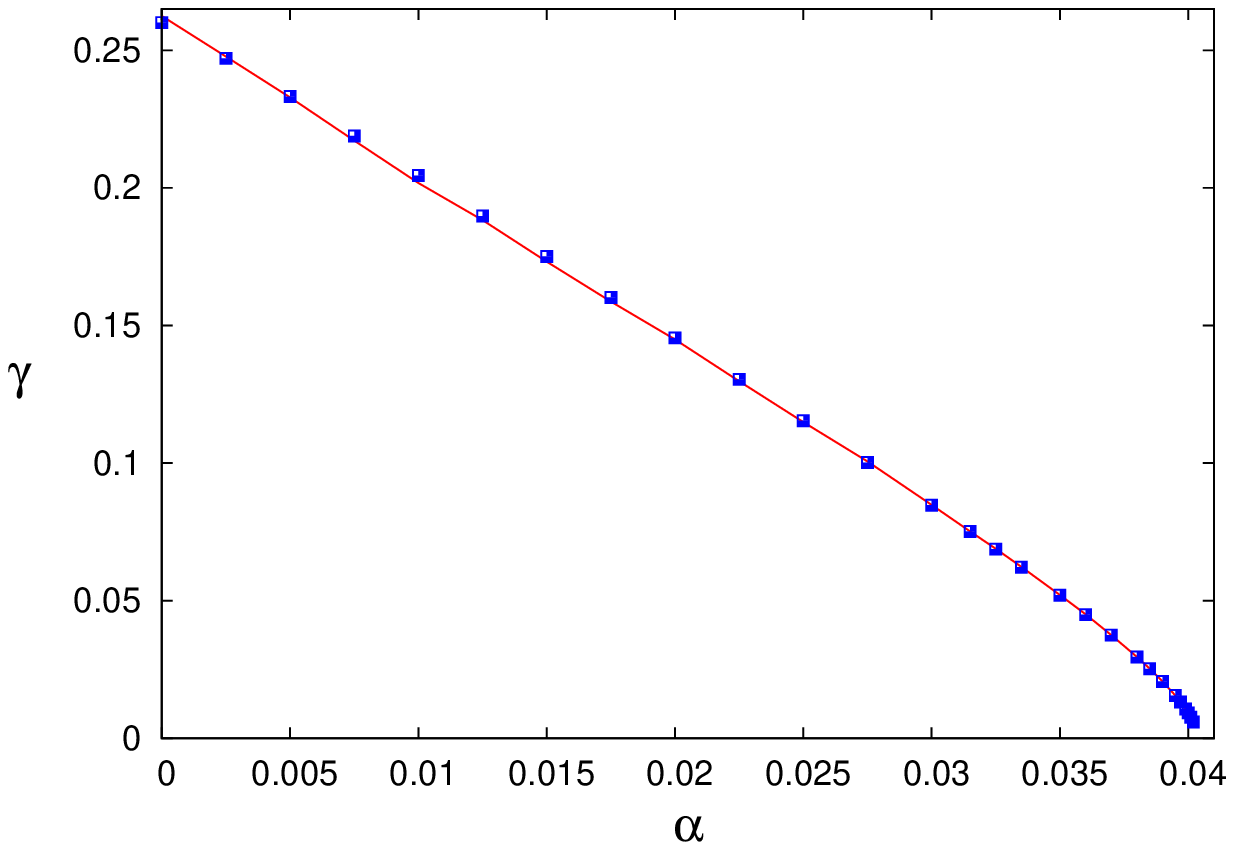}
\end{tabular}
\caption{\small{The dependence of the parameters of the fundamental quasinormal mode on the
coupling constant. The solid lines and points denote data obtained from fit to numerical
solutions and by means of shooting technique, respectively.}} \label{fig5}
\end{figure}

Values of the QNR parameters as a function of dimensionless coupling constant $\alpha$ are
plotted in Fig.\ref{fig5}. Both parameters decrease with growing $\alpha$ and go to zero as
$\alpha$ approaches the critical value  $\alpha_{crit}$. It means that, as $\alpha \rightarrow
\alpha_{crit}$ the wavelength of perturbation of Skyrmion and life-time of these perturbations
are growing to $\infty$.
\newpage

\begin{figure}[htbp]
\begin{center}
\includegraphics[width=0.45\textwidth]{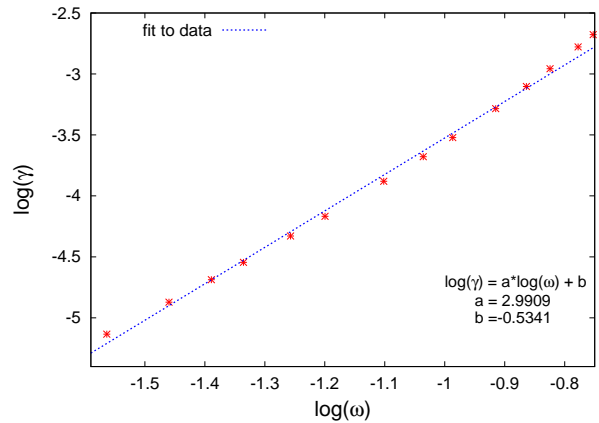}
\caption{The dispersion relation at the critical value of coupling constant
 $\alpha \rightarrow \alpha_{crit}$ }
\label{fig6}
\end{center}
\end{figure}

In Fig.\ref{fig6} we present the dispersion relation  $\gamma$ vs. $\omega$  for $\alpha \rightarrow \alpha_{crit}$. Numerical analysis seems to suggest, that this relations takes the form: $\gamma \sim \omega^3$.

We have also studied late time asymptotics, where quasi-normal modes are negligible and the
relaxation process is dominated by power-law tail. 
Our numerical calculations show that at late times the relaxation to the Skyrmion takes the form: $F(t,r) - S(r) \sim t^{- \gamma} $ where the value of $\gamma$ depends on the coupling constant $\alpha$ and is equal 5   for $\alpha=0$ and 4 for $\alpha > 0$. As it was already pointed out in \cite{BChR1} for the case of flat space, this does not agree with the predictions of linear scattering theory (see e.g. \cite{Price,Ching}).
According to this theory, for late times the evolution of the system is well described by a
linear wave equation with potential $V(r)$ (see Eq.(\ref{pulsation_v})). 
Let us consider more general case, described by the equation:
\begin{equation}\label{lin_tail}
   -\Psi''+ \left(\frac{l (l+1)}{r^2} + V \right) \Psi = k^2 \Psi,
\end{equation}
where $V(r)$ is the finite part of potential.
 If the fall-off of $V(r)$ at spatial infinity is $\beta$, i.e. $V(r) \sim r^{-\beta}$ 
for  $r  \rightarrow \infty$, than for compactly supported initial data linear theory 
predicts: $\gamma = 2l + \beta$.
As in our problem we have $l=1$ and $\beta=6$ for flat and $\beta=3$ for gravitating case (see
Section III), we may expect 8 and 5 for the values of power-law tail exponent for flat and
gravitating model respectively, which is in disagreement with numerical results. This is  another
example of a situation where the tail has genuinely non-linear character and linear theory fails.
To describe the tail correctly, one should use a nonlinear perturbatively scheme proposed
recently by Bizon et. al. \cite{BChR1,BCR2,BCR3,BCR4,BCR5}. The application of this technique to
the case of $B=1$ sector of ES model is tedious, so we have checked, that we get the same values
of tail exponents in the case of $B=0$ sector of ES model and in non-linear sigma model. Analyzes of tails in these models will be published elsewhere (see \cite{BChRS} for more details).

\vskip 0.2cm \noindent \textbf{Acknowledgments:} I would like to thank Tadeusz Chmaj for suggesting this subject and many useful hints and discussions. This research was supported in part by MNII grant NN202 079235.

\end{document}